\documentclass[twoside]{article}
\usepackage{fleqn,espcrc2}

\usepackage{graphicx}

\newcommand{\nn}{\nonumber\\}
\newcommand{\qmul}{\mbox
{$*$ \hskip -0.80em \raisebox{-0.8ex}{$\scriptstyle \epsilon$}}}
\newcommand{\qodd}{\mbox
{$*$ \hskip -0.75em \raisebox{-0.9ex}{$\scriptscriptstyle 1$}}}
\newcommand{\qeven}{\mbox
{$*$ \hskip -0.75em \raisebox{-0.9ex}{$\scriptscriptstyle 0$}}}
\newcommand{\lpar}{\stackrel{\leftarrow}{\partial}}
\newcommand{\rpar}{\stackrel{\rightarrow}{\partial}}

\newcommand{\AmS}{{\protect\the\textfont2
  A\kern-.1667em\lower.5ex\hbox{M}\kern-.125emS}}

\title{Supersymmetry and the Odd Poisson Bracket}

\author{V.A. Soroka\address{Institute for Theoretical Physics, NSC
	"Kharkov Institute of Physics and Technology",\\
	61108 Kharkov, Ukraine}
	\thanks{This work was supported in part by the Ukrainian State
	Foundation of Fundamental Researches, Grant No 2.5.1/54.}}

\begin{document}

\begin{abstract}
Some applications of the odd Poisson bracket developed by Kharkov's
theorists are represented.
\vspace{1pc}
\end{abstract}

\maketitle

\section{INTRODUCTION}

The theory of supersymmetry substantially expanded our ideas not only on
the possible types of symmetry relations but also on the way of the
dynamics description. Indeed, with the discovery of supersymmetry
\cite{gl,va,vs,wz}, which extends to a superspace the configuration space
$x$ by adding to it anticommuting (Grassmann) coordinates $\theta$ and
extends by this also the phase space with the coordinates $x$ and $p$, the
possibility arises to generalize directly the Poisson bracket to the even
Poisson bracket by adding the Martin bracket \cite{mar} which canonically
conjugates the Grassmann variables $\theta$ with each other. However,
mathematicians (first of them was Buttin \cite{but}) proved that this is
not the only possibility of generalization of the Poisson bracket to the
case of the phase superspace having both the even and odd (with respect
to the Grassmann parity) variables. In addition to the even
Poisson--Martin bracket indicated above there exists another possibility
to define on the phase superspace the bracket operation of the Poisson
type which canonically conjugates phase variables of the opposite
Grassmann parity.  These two brackets confine all possible cases of
combinations of the Grassmann parity of canonically conjugated quantities
in the superspace.  The essential difference of the latter bracket from
the even bracket is the existence in it of a nontrivial Grassmann parity
equal to unity, and because of this it is called the odd Poisson bracket
or simply an odd bracket (OB).

The even Poisson and Martin brackets are well known. Their quantization
leads to respectively two forms of quantum statistics: the Bose--Einstein
and the Fermi--Dirac. The study and physical use of the OB has started
relatively recent. In physics the OB has firstly appeared as an adequate
language for the quantization of the gauge theories in the well--known
Batalin--Vilkovisky scheme \cite{bv}. Leites \cite{l} assumed that the OB
is connected with another type of classical mechanics. Manin and Radul
have used the OB for the description of the supersymmetric integrable
systems \cite{mr} and Kupershmidt has applied it for the description of
the one--dimensional ideal liquid with fermionic degrees of freedom
\cite{k}.

However, the apparent dynamical role of the OB was not understood quite
well till papers \cite{vpst,s}. In \cite{vpst} a possibility of the
reformulation of Hamiltonian dynamics on the basis of the OB with the use
of the dynamically equivalent Grassmann--odd Hamiltonian was proved for
the classical supersymmetric Witten mechanics having two even $q$, $p$ and
two odd $\theta_1$, $\theta_2$ phase variables and in \cite{s} this result
was generalized for the Hamilton systems with an arbitrary equal number of
pairs of even and odd (relative to the Grassmann grading) phase
coordinates. Nevertheless, there are such Hamiltonian systems which have a
description only by means of the OB and can not be reformulated in terms
of the even bracket. In particular, as it was shown in \cite{vty}, usual
hydrodynamics is just such a system with three even and three odd phase
coordinates. Hamiltonian formulation of dynamics in the OB is closely
related with Lagrangian formulation having a Grassmann--odd Lagrangian. As
an example of this in \cite{s1} an action with the Grassmann--odd
Lagrangian for the supersymmetric classical Witten mechanics has been
constructed.

The direct approach to the quantization of the OB leads to the odd Planck
constant which has no physical interpretation. Therefore in \cite{vst}
another, free from this difficulty, prescription for the canonical
quantization of the OB was suggested and several odd--bracket quantum
representations for the canonical variables were also obtained. In
contrast with the even bracket case, some of the odd--bracket quantum
representations turn out to be nonequivalent \cite{vs1,s2}. The direct
connection of the odd--bracket quantum representations with the
quantization of classical Hamiltonian dynamics based on the OB has been
demonstrated on an example of the one--dimensional supersymmetric
oscillator \cite{s2}. Another application of the odd--bracket quantum
representations is connected with the possibility of realization of the
idea of the composite (spinor) structure of coordinates, which illustrated
on the examples of superspaces of the orthosymplectic supergroups
$OSp(N,2k)$ for $k=1,2$ \cite{vst,v}.

Recently a linear OB on Grassmann algebra has been introduced \cite{s3}.
It was revealed that with the OB, which corresponds to a semi--simple Lie
group, both a Grassmann--odd Casimir function and invariant (relative to
this group) nilpotent differential operators of the first, second and
third order with respect to Grassmann derivatives are naturally related
and enter into a finite--dimensional Lie superalgebra. A connection of the
quantities, forming this superalgebra, with the BRST charge,
$\Delta$--operator and ghost number operator was indicated.

\section{CLASSICAL ASPECT OF THE ODD BRACKET}

\subsection{Properties of the odd Poisson bracket}

First, we recall the necessary properties of various graded
Poisson brackets. The even and odd brackets in terms of the real even
${y_i} = (q^a,p_a)$ and odd $\eta ^i = \theta ^{\alpha }$  canonical
variables have, respectively, the form
\begin{eqnarray}\label{1}
\{ A , B \}_o = A [\sum_{a = 1}^n
\left(\stackrel{\leftarrow}{\partial}_{q^a}
\stackrel{\rightarrow}{\partial}_{p_a} -
\stackrel{\leftarrow}{\partial}_{p_a}
\stackrel{\rightarrow}{\partial}_{q^a}\right)\nn -
i \sum_{\alpha =1}^{2n}\stackrel{\leftarrow}{\partial}_{\theta ^{\alpha
}} \stackrel{\rightarrow}{\partial}_{\theta^{\alpha }}
] B\ ;
\end{eqnarray}
\begin{equation}\label{2}
\{ A , B \}_1 = A \sum^{N}_{ i=1}
\left(\stackrel{\leftarrow}{\partial}_{y_i}
\stackrel{\rightarrow}{\partial}_{\eta ^i} -
\stackrel{\leftarrow}{\partial}_{\eta ^i}
\stackrel{\rightarrow}{\partial}_{y_i}\right) B\ ,
\end{equation}
where $\stackrel{\leftarrow}{\partial}$ and $\stackrel{\rightarrow}{\partial}$
are the right and left derivatives, and the notation $\partial_x =
{\partial \over {\partial x}}$ is introduced.  By introducing apart from
the Grassmann grading $g(A)$ of any quantity $A$ its corresponding bracket
grading $g_{\epsilon }(A) = g(A) + \epsilon \pmod 2$ ($\epsilon  = 0,1$),
the grading and symmetry properties, the Jacobi identities and the
Leibnitz rule are uniformly expressed for the both brackets (1,2) as
\begin{equation}\label{3a}
g_\epsilon(\{A _{,} B\}_\epsilon) = g_\epsilon(A) + g_\epsilon(B)\pmod 2\ ,
\end{equation}
\begin{equation}\label{3b}
\{ A , B \}_\epsilon = -(-1) ^{g_\epsilon (A)g_\epsilon (B)}\
\{ B , A \}_\epsilon\ ,
\end{equation}
\begin{equation}\label{3c}
\sum_{(ABC)}(-1) ^{g_\epsilon (A)g_\epsilon (C)}\
\{ A , \{ B , C \}_\epsilon\}_\epsilon = 0\ ,
\end{equation}
\begin{eqnarray}\label{3d}
\{ A , B C \}_{\epsilon } = \{ A , B \}_{\epsilon }\ C\nn +
(- 1) ^{g_\epsilon (A)g(B) }\ B \{ A , C \}_{\epsilon }\ ,
\end{eqnarray}
where (\ref{3a})--(\ref{3c}) have the shape of the Lie superalgebra
relations in their canonical form \cite{ber1} with $g_\epsilon(A)$ being
the canonical grading for the corresponding bracket.

In terms of arbitrary real dynamical variables $x^M = (x^m, x^\alpha) =
x^M(y,\eta)$ with the same number of Grassmann even $x^m$ and odd
$x^\alpha$ coordinates the odd bracket (2) takes the form
\begin{equation}\label{4}
\{ A , B \}_1 = A \stackrel{\leftarrow}{\partial}_M \bar \omega^{MN}(x)
\stackrel{\rightarrow}{\partial}_N B\ .
\end{equation}
The matrix $\bar \omega_{MN}$, inverse to $\bar \omega^{MN}$
\begin{equation}\label{5}
\bar \omega_{MN}\bar \omega^{NL} = \delta_M^L\ ,
\end{equation}
and consisting of the coefficients of the odd closed 2-form, in view of
the odd bracket properties (\ref{3a})-(\ref{3c}) can be represented in the
form of the grading strength
\begin{equation}\label{6}
\bar \omega_{MN} = \partial_M \bar{\cal{A}}_N - (-1)^{g(M)g(N)} \partial_N
\bar {\cal{A}}_M\ ,
\end{equation}
where $g(M) = g(x^M)$ and $\partial_M
= \partial  / \partial x^M$. The coefficients of the 1-form $\bar
{\cal{A}}(d) = dx^M\bar {\cal{A}}_M$ satisfy the conditions
\begin{equation}\label{7}
g(\bar {\cal{A}}_M) = g(M) + 1\ ,\qquad (\bar {\cal{A}}_M)^+ =
\bar {\cal{A}}_M\  .
\end{equation}
As can be seen from (\ref{6}) $\bar \omega_{MN}$ is invariant under
gauge transformations
\begin{equation}\label{8}
\bar {\cal{A}}'_M = \bar {\cal{A}}_M +
\partial_M \bar \chi
\end{equation}
with functions $\bar \chi$ as parameters.

\subsection{Classical dynamics in terms of the odd Poisson bracket}

Let us consider the Hamilton system containing an equal number $n$ of
pairs of even and odd with respect to the Grassmann grading real canonical
variables. We require that the equations of motion of the system be
reproduced both in the even Poisson-Martin bracket (1) with the help of the
even Hamiltonian $H$ and in the odd bracket (2) with the Grassmann-odd
Hamiltonian $\bar H$, that is \cite{vpst,s},
\begin{equation}\label{9}
{dx^M\over dt} = \{x^M, H\}_0 = \{ x^M, \bar H \}_{1}\ ,
\end{equation}
where t is the proper time. Using definitions (\ref{4}) and (1) together
with (\ref{5}), (\ref{6}) the relations (\ref{9}) can be represented as the
equations
\begin{eqnarray}\label{10}
(\partial_M \bar{\cal{A}}_N - (-1)^{g(M)g(N)} \partial_N \bar{\cal{A}}_M)
\omega^{NL} {\partial_L} H =\nonumber\\
{\partial_M} \bar H
\end{eqnarray}
to derive the unknown $\bar H$ and $\bar{\cal{A}}_M$ under the given $H$
and the even matrix $\omega^{MN}$ corresponding to the even bracket (1).

In order to solve equations (\ref{10}) it is convenient to use such real
canonical in the even bracket (1) coordinates $x^M$ which contain among
canonically conjugate pairs the pair consisting of the proper time $t$ and
the Hamiltonian $H$. It follows from (\ref{9}) that the rest of the
canonical quantities $z^{M}$ would be the integrals of motion for the
system considered: even $I_1,...,I_{2(n-1)}$ and odd
$\Theta^1,...,\Theta^{2n}$.  In terms of these coordinates $x^{M}$
equations (\ref{10}) take the form
\begin{equation}\label{11}
(\partial_M \bar{\cal{A}}_t - \partial_t \bar{\cal{A}}_M)=
{\partial_M} \bar H\ .
\end{equation}
The quantities $\bar{\cal{A}}_M, \bar \chi$ and $\bar H$ can be
expanded in powers of the Grassmann variables $\Theta^{\alpha }$ as
\begin{eqnarray}\label{12a}
\bar{\cal{A}}_m =\nonumber\\
\sum_{k=1}^{n} {i^{(k-1)(2k-1)} \over (2k-1)!}
A_{m\alpha_1\dots \alpha_{2k-1}}\Theta^{\alpha_1 }
\dots\Theta^{\alpha_{2k-1} }\ ,
\end{eqnarray}
\begin{equation}\label{12b}
\bar{\cal{A}}_\alpha = \sum_{k=0}^{n} {i^{k(2k+1)} \over (2k)!}
B_{\alpha \alpha_1\dots \alpha_{2k}}\Theta^{\alpha_1 }
\dots\Theta^{\alpha_{2k} }\ ,
\end{equation}
\begin{eqnarray}\label{12c}
\bar \chi =\nonumber\\
\sum_{k=1}^{n} {i^{(k-1)(2k-1)} \over (2k-1)!}
\chi_{\alpha_1\dots \alpha_{2k-1}}\Theta^{\alpha_1 }
\dots\Theta^{\alpha_{2k-1} }\ ,
\end{eqnarray}
\begin{eqnarray}\label{12d}
\bar H = \nonumber\\
\sum_{k=1}^{n} {i^{(k-1)(2k-1)} \over (2k-1)!}
h_{\alpha_1\dots \alpha_{2k-1}}\Theta^{\alpha_1 }
\dots\Theta^{\alpha_{2k-1} }\ .
\end{eqnarray}
The $\Theta^{\alpha }$ coefficients are the real Grassmann-even functions
of the even variables $x^m = (t,H,I_1,\dots,I_{2(n-1)})$ and are chosen to
be antisymmetric in the indices contracted with $\Theta^{\alpha
}$. In terms of these functions the gauge transformations (\ref{8}) have
the form
\begin{eqnarray}
A'_{m\alpha_1\dots \alpha_{2k-1}} = A_{m\alpha_1\dots \alpha_{2k-1}} +
\nonumber\\
\partial_m  \chi_{\alpha_1\dots \alpha_{2k-1}}\ , (k = 1,\dots,n)\ ;
\nonumber
\end{eqnarray}
\begin{eqnarray}
B'_{[\alpha \alpha_1\dots \alpha_{2k}]} = B_{[\alpha \alpha_1\dots
\alpha_{2k}]} +\nonumber\\
\chi_{\alpha\alpha_1\dots \alpha_{2k}}\ ,
(k = 0,1,\dots,n-1)\ ;\nonumber
\end{eqnarray}
\begin{eqnarray}\label{13}
B'_{(\alpha \alpha_j) \alpha_1\dots \alpha_{j-1}\alpha_{j+1}\dots \alpha_{2k}}
=\nonumber\\
B_{(\alpha \alpha_j) \alpha_1\dots \alpha_{j-1}\alpha_{j+1}\dots
\alpha_{2k}}\ , \left(k = 0,1,\dots,n \atop j = 1,\dots,2k \right)\ ;
\end{eqnarray}
where the expansion in the components with different symmetries of the
indices has been used for the tensor antisymmetric in all indices but the
first
\begin{eqnarray}
B_{\alpha \alpha_1\dots \alpha_{2k}} = B_{[\alpha \alpha_1\dots
\alpha_{2k}]} +\nonumber\\
{2 \over {N + 1}} \sum_{j=1}^{N} (-1)^{j-1}
B_{(\alpha \alpha_j) \alpha_1\dots \alpha_{j-1}\alpha_{j+1}\dots
\alpha_{2k}}\ . \nonumber
\end{eqnarray}

The additive character of the transformations for the functions
$B_{[\alpha \alpha_1\dots \alpha_{2k}]}$ $(k = 0,1,\dots,n-1)$ allows us to
put them equal to zero in the expression (\ref{12b}) for
$\bar{\cal{A}}_\alpha$ by choosing $\chi_{\alpha\alpha_1\dots \alpha_{2k}}
= - B_{[\alpha \alpha_1\dots \alpha_{2k}]}$. This gauge choice amounts to
the following gauge condition
\begin{eqnarray}
\Theta^{\alpha } \bar{\cal{A}}_\alpha = 0\ .\nonumber
\end{eqnarray}
Using this condition and equations (\ref{11}), we obtain the equality
\begin{equation}\label{14}
\bar H = \bar{\cal{A}}_t\ .
\end{equation}
which, being substituted again into (\ref{11}), leads to the simple
equations
\begin{equation}\label{15}
{\partial}_t \bar{\cal{A}}_M = 0\ .
\end{equation}
Thus, in consequence of (\ref{15}), the solution of equations (\ref{11})
for $\bar{\cal{A}}_M$ and $\bar H$ in the chosen gauge resides in that the
nonzero coefficients $A_{m\alpha_1\dots \alpha_{2k-1}}$ and $B_{(\alpha
\alpha_j) \alpha_1\dots \alpha_{j-1}\alpha_{j+1}\dots \alpha_{2k}}$ in
expansions (\ref{12a},\ref{12b}) for $\bar{\cal{A}}_M$ are the arbitrary
functions (denoted as $a_{m\alpha_1\dots \alpha_{2k-1}a_{m\alpha_1\dots
\alpha_{2k-1}}}$ and $b_{(\alpha \alpha_j) \alpha_1\dots
\alpha_{j-1}\alpha_{j+1}\dots \alpha_{2k}}$, respectively) of all, except
the proper time $t$, even variables  $H$ and $I_1,\dots,I_{2(n-1)}$, and
the odd Hamiltonian is expressed in terms of these functions with the help
of equation (\ref{14}).

Using the gauge transformations (\ref{13}) with the arbitrary functions
$\chi_{\alpha_1\dots \alpha_{2k-1}}(t,H,I)$, we obtain the general
solution of equations (\ref{11}) in the arbitrary gauge:
\begin{eqnarray}
A_{m\alpha_1\dots \alpha_{2k-1}} =
a_{m\alpha_1\dots \alpha_{2k-1}}(H,I) +\nonumber\\
\partial_m \chi_{\alpha_1\dots \alpha_{2k-1}}(t,H,I)\ ;\nonumber
\end{eqnarray}
\begin{eqnarray}
B_{\alpha \alpha_1\dots \alpha_{2k}} =\nonumber\\
{2 \over {2k + 1}} \sum_{j=1}^{2k} (-1)^{j-1}
b_{(\alpha \alpha_j) \alpha_1\dots \alpha_{j-1}\alpha_{j+1}\dots
\alpha_{2k}}(H,I) +\nonumber\\
\chi_{\alpha\alpha_1\dots \alpha_{2k}}(t,H,I)\ ;\nonumber
\end{eqnarray}
\begin{eqnarray}
h_{\alpha_1\dots \alpha_{2k-1}} =
a_{t\alpha_1\dots \alpha_{2k-1}}(H,I)\ .\nonumber
\end{eqnarray}
Note that the solution of the analogous problem of finding the even
brackets and the corresponding even Hamiltonians, which lead to the same
equations of motion
\begin{eqnarray}
{dx^M\over dt} = \{x^M, H\}_0 = \{ x^M, \tilde H \}_{\tilde 0}\ ,
\nonumber
\end{eqnarray}
has a similar structure but with the difference that the odd quantities
$\bar{\cal{A}}_M, \bar \chi$ and $\bar H$ has to be replaced by the even
ones.

Thus, we extended the notion of the bi-Hamiltonian systems onto the case
when the pairs of the Hamiltonian-bracket, giving the same equations of
motion, have an opposite Grassmann grading.

\subsection{Hydrodynamics description in the odd bracket}

In \cite{vty} an approach for the description of hydrodynamics as a
Hamilton system in terms of the odd bracket has been developed. This
approach is a powerful tool for the description and construction of
hydrodynamic invariants.

The velocity field for the hydrodynamic medium with coordinates $x^i(t)$
satisfies the equations
\begin{eqnarray}
{dx^i\over dt} = V^i(x,t)\ ,
\nonumber
\end{eqnarray}
which can be rewritten
\begin{eqnarray}
{dx^i\over dt} = \{x^i, H\}_1 = V^i(x,t)
\nonumber
\end{eqnarray}
by means of the odd bracket (2)
\begin{eqnarray}
\{ A , B \}_1 = A \sum^{3}_{ i=1}
\left(\stackrel{\leftarrow}{\partial}_{x^i}
\stackrel{\rightarrow}{\partial}_{\theta _i} -
\stackrel{\leftarrow}{\partial}_{\theta _i}
\stackrel{\rightarrow}{\partial}_{x^i}\right) B
\nonumber
\end{eqnarray}
with the help of the Grassmann--odd Hamiltonian
\begin{eqnarray}
H=V^i(x,t)\theta_i\ .
\nonumber
\end{eqnarray}

For the concordance of the equations of motion for Grassmann coordinates
$\theta_i$
\begin{equation}\label{a1}
{d\theta_i\over dt} = \{\theta_i, H\}_1 = - \partial_{x^i}V^k\theta_k
\end{equation}
with the equations for the exterior differentials $dx^i$
\begin{equation}\label{a2}
{d\over dt}dx^i = dx^k \partial_{x^k}V^i
\end{equation}
let us introduce a Lie dragged metric tensor $g_{ik}(x,t)$
\begin{eqnarray}
\partial_tg_{ik}+L_Vg_{ik}=0\ ,
\nonumber
\end{eqnarray}
($L_V$ is the Lie derivative) which connects the field co--differentials
$\theta_i$ with differentials $dx^i$
\begin{eqnarray}
\theta_i=g_{ik}dx^k\ .
\nonumber
\end{eqnarray}

A super invariant $I_s$ is a function of the superspace coordinates
$z=(x^i, \theta_i)$ $(i=1,2,3)$
\begin{equation}\label{a3}
I_s(z,t)=I+J^i\theta_i+S_i{\epsilon^{ikl}\over\rho}\theta_k\theta_l+
{1\over\rho}\theta_1\theta_2\theta_3
\end{equation}
and obeys the equation for an invariant
\begin{equation}\label{a4}
{dI_s\over dt} = \partial_tI_s+\{I_s, H\}_1 = 0\ .
\end{equation}
The $\theta_i$ coefficients of the super invariant (\ref{a3}) are
well--known local hydrodynamic invariants: $I(x,t)$ is the Lagrangian
invariant, $J^i(x,t)$ is a frozen-in field, $S_i(x,t)$ is a frozen-in
surface and $\rho(x,t)$ is the invariant density. The $\theta_i$ components
of the super equation (\ref{a4}) are the equations for the local invariants
\begin{equation}\label{a5}
\partial_tI+L_VI=0\ ,
\end{equation}
\begin{equation}\label{a6}
\partial_tJ^i+L_VJ^i=0\ ,
\end{equation}
\begin{equation}\label{a7}
\partial_tS_i+L_VS_i=0\ ,
\end{equation}
\begin{equation}\label{a8}
\partial_t\rho+\partial_{x^i}(V^i\rho)=0\ .
\end{equation}
From two super invariants $I_{s1}$ and $I_{s2}$ we can construct new ones
with the use of the usual multiplication
\begin{eqnarray}
I_{s3}=I_{s1}I_{s2}
\nonumber
\end{eqnarray}
and with the help of the odd bracket composition
\begin{eqnarray}
I_{s4}=\{I_{s1}, I_{s2}\}\ .
\nonumber
\end{eqnarray}
In consequence of the equations (\ref{a1}), (\ref{a2}) and
(\ref{a5})--(\ref{a8}) all terms of the super invariant (\ref{a3}) are
separately conserved as well as the quantity
\begin{equation}\label{a9}
{d\over dt}(\theta_idx^i) =0\ .
\end{equation}
Due to (\ref{a9}) we can pass from local invariants to integral ones by
taking a Fourier transform of the super invariant (\ref{a3})
\begin{equation}\label{a10}
\int I_s(z,t) \rho(x,t) \exp(\theta_idx^i)d\theta^1d\theta^2d\theta^3=
I_{\rm int}\ .
\end{equation}
By performing in (\ref{a10}) the Berezin integration \cite{ber1} over
Grassmann variables $\theta_i$ we obtain as a result the ordinary integral
conservation laws: "mass" conservation law, frozen-in field flux
conservation law and conservation of the circulation
${\bf S}$-invariant respectively
\begin{eqnarray}
I_{\rm int}=\int\rho Idx^1\wedge dx^2\wedge dx^3+
\int\rho{\bf J}d{\bf x}\wedge{\bf x}+\nonumber\\
\int S_idx^i+1
\nonumber
\end{eqnarray}

\subsection{Dynamics with Grassmann--odd Lagrangian}

Here we show that the idea
stated in \cite{s} about the possible
existence of the dynamics formulation with the Grassmann-odd Lagrangian
can be realized on the example of $d = 1, N = 2$ supersymmetric Witten's
mechanics \cite{w} in its classical version \cite{vpst,ap}.

Let us consider a system invariant with respect to the $N = 2\
(\alpha = 1,2)$ supersymmetry of the proper time $t$
\begin{eqnarray}
t' = t + i\epsilon^\alpha \theta^\alpha\ , \qquad
\theta'^\alpha\ = \theta^\alpha + \epsilon^\alpha\ ,
\nonumber
\end{eqnarray}
\begin{eqnarray}
D_\alpha = {\partial \over \partial \theta^\alpha} - i
\delta_{\alpha\beta}\theta^\beta {\partial \over \partial t}\ .
\nonumber
\end{eqnarray}
By using the covariant derivatives $D_\alpha$
and two real scalar superfields
\begin{eqnarray}
\Phi(t,\theta^\alpha) = q(t) +
i {\psi_\alpha}(t) \theta^\alpha + i F(t) \theta^\alpha\theta_\alpha \ ,
\nonumber
\end{eqnarray}
\begin{eqnarray}
\Psi(t,\theta^\alpha) = \eta(t) + i {a_\alpha}(t) \theta^\alpha
 + i \Xi(t) \theta^\alpha\theta_\alpha       \ ,
\nonumber
\end{eqnarray}
having the opposite values of the Grassmann grading $g$ $(g(\Phi) = 0,
g(\Psi) = 1)$, the following supersymmetric action $\bar S$ with the
Grassmann-odd Lagrangian $\bar L$ $(g(\bar L) = 1)$ can be constructed
\begin{eqnarray}\label{b1}
\bar S = \int dtd\theta_2d\theta_1
\left[ - {1 \over2} D^\alpha \Psi D_ \alpha \Phi + i \Psi W(\Phi)\right]
\nn
= \int dt \bar L\ ,
\end{eqnarray}
where $W(\Phi)$ is an arbitrary real function
of $\Phi$.
Excluding in (\ref{b1}) the auxiliary fields $F$ and $\Xi$, we obtain
\begin{eqnarray}\label{b2}
\bar L =\ \stackrel{\cdot}{\eta} \stackrel{\cdot}{q} +
{i \over 2} ( a^\alpha\stackrel{\cdot}{\psi^\alpha} -
\stackrel{\cdot}{a^\alpha}\psi^\alpha - 2 a^\alpha \psi_\alpha W')\nn
-\eta\ ( WW' + {i \over 2} \psi^\alpha \psi_\alpha W'')\ ,
\end{eqnarray}
where the dot and the prime mean the derivatives with respect to
$t$ and $q$ correspondingly. The odd Lagrangian $\bar L$ leads to
the momenta $p, \pi, \pi^\alpha, p^\alpha$
canonically conjugate in the odd bracket to the coordinates
$\eta$, $q$, $a^\alpha$ and $\psi^\alpha$
\begin{eqnarray}
\{\eta, p\}_1 = \{q, \pi \}_1 = 1;
\nonumber
\end{eqnarray}
\begin{eqnarray}
\{a^\alpha, \pi^\beta \}_1 = \{\psi^\alpha, p^\beta \}_1 =
\delta^{\alpha \beta}\ .
\nonumber
\end{eqnarray}
There are the second-class constraints
$\varphi^\alpha = \pi^\alpha + {i
\over 2} \psi^\alpha,\  f^\alpha = - p^\alpha + {i \over 2}
a^\alpha$
\begin{eqnarray}
\{ \varphi^\alpha, f^\beta
\}_1 = - i \delta^{\alpha \beta} ;
\nonumber
\end{eqnarray}
\begin{equation}\label{b3}
\{ \varphi^\alpha, \varphi^\beta
\}_1 = \{ f^\alpha, f^\beta \}_1 = 0\ .
\end{equation}
In terms of the variables
$\chi^\alpha = \pi^\alpha - {i \over 2}
\psi^\alpha,\ g^\alpha = - p^\alpha - {i \over 2} a^\alpha$
Dirac's odd bracket from any
functions $A$ and $B$ takes the form
\begin{eqnarray}\label{b4}
\{A, B\}_1^{D.B.} =
A ({\lpar}_q{\rpar}_\pi - {\lpar}_\pi{\rpar}_q +
{\lpar}_\eta{\rpar}_p \nn
- {\lpar}_p{\rpar}_\eta +
i {\lpar}_{\chi^\alpha}{\rpar}_{g^\alpha} -
i {\lpar}_{g^\alpha}{\rpar}_{\chi^\alpha} ) B\ ,
\end{eqnarray}
The total odd Hamiltonian following from (\ref{b2}), if subjected
to the second-class constraints $\varphi^\alpha = 0$ and
$f^\alpha = 0$, takes the form
\begin{eqnarray}\label{b5}
\bar H = p\pi + \eta\ (WW' + {i
\over 2} \chi_\alpha \chi^\alpha W'')\nn
+ i g_\alpha\chi^\alpha W'
\end{eqnarray}
and in Dirac's bracket (\ref{b4}) gives Hamilton's equations
$\stackrel{\cdot}{z^a} = \{z^a, \bar H \}_1^{D.B.}$
for the independent phase variables $z^a = (q, p, \chi^\alpha;\
\eta, \pi, g^\alpha)$
\begin{eqnarray}
\stackrel{\cdot}{q} = p\ ,
\nonumber
\end{eqnarray}
\begin{eqnarray}
\stackrel{\cdot}{p} = -\ WW' - {i \over 2} \chi_\alpha \chi^\alpha W''\ ,
\nonumber
\end{eqnarray}
\begin{eqnarray}\label{b6a}
\stackrel{\cdot}{\chi^\alpha} = \chi_\alpha W'\ ;
\end{eqnarray}
\begin{eqnarray}
\stackrel{\cdot}{\eta} = \pi\ ,
\nonumber
\end{eqnarray}
\begin{eqnarray}
\stackrel{\cdot}{\pi} = -\ \left\{\ {\eta \over 2}\ [(W^2)'' +
\chi_\alpha \chi^\alpha W'''] + i g_\alpha \chi^\alpha W''\right\}\ ,
\nonumber
\end{eqnarray}
\begin{eqnarray}\label{b6b}
\stackrel{\cdot}{g^\alpha} = g_\alpha W' + \eta \chi_\alpha W''\ .
\end{eqnarray}
Equations (\ref{b6a}) are Hamilton's equations for Witten's
supersymmetric mechanics \cite{w} in its classical version \cite{vpst}
which can be derived by means of Dirac's even bracket
\begin{eqnarray}\label{b7}
\{A, B\}_0^{D.B.} =
\{A, B\}_0 - i \{A, \varphi^\alpha\}_0\{\varphi^\alpha, B\}_0\nn
=A \left({\lpar}_q{\rpar}_p - {\lpar}_p{\rpar}_q +
i {\lpar}_{\chi^\alpha}{\rpar}_{\chi^\alpha} \right) B
\end{eqnarray}
with the help of the even Hamiltonian $H$
\begin{equation}\label{b8}
H = {{p^2 + W^2(q)}\over 2} + {i \over 2} \chi_\alpha \chi^\alpha W'(q)\ ,
\end{equation}
which both follow from the $N = 2$ supersymmetric action
with the Grassmann-even Lagrangian L $(g(L) = 0)$ (see, for example,
\cite{ap})
\begin{eqnarray}
S = {1 \over 4} \int dtd\theta_2d\theta_1
\left[ D^\alpha \Phi D_ \alpha \Phi + 2i V(\Phi)\right]\nn
=\int dt  L\ ,
\nonumber
\end{eqnarray}
where $V'(\Phi) = 2W(\Phi)$ and the even Lagrangian after exclusion of the
auxiliary field $F$ is
\begin{equation}\label{b9}
L = {1\over 2}\ [ \stackrel{\cdot}{q}^2
+\ i ( \psi^\alpha\stackrel{\cdot}{\psi^\alpha}
+\ \psi_\alpha \psi^\alpha W') - W^2 ]\ .
\end{equation}
The momenta canonically conjugate in the even bracket to the coordinates
$q$ and $\psi^\alpha$
\begin{eqnarray}
\{q, p\}_0 = 1; \qquad \{\psi^\alpha, \pi^\beta \}_0 =
- \delta^{\alpha \beta}\ ,
\nonumber
\end{eqnarray}
following from the even Lagrangian (\ref{b9}),
lead to the second-class constraints
\begin{equation}\label{b10}
\varphi^\alpha = \pi^\alpha + {i \over 2} \psi^\alpha\ ;\qquad
\{ \varphi^\alpha, \varphi^\beta \}_0 = - i
\delta^{\alpha \beta}\ ,
\end{equation}
which commute in the even bracket with the variables
$\chi^\alpha = \pi^\alpha - {i \over 2} \psi^\alpha$,
entering into the definitions for the even Dirac bracket (\ref{b7}) and
the even Hamiltonian (\ref{b8}). The even Hamiltonian (\ref{b8}) follows
with the use of the second-class constraints restriction $\varphi^\alpha =
0$ from the total even Hamiltonian corresponding to the Lagrangian
(\ref{b9}).

Equation (\ref{b6b}) can be obtained by taking
the exterior differential $d$ of the Hamilton equations (\ref{b6a}) for the
Witten mechanics and performing the map $\lambda$:
\begin{eqnarray}
dq \rightarrow \eta ,\  dp \rightarrow \pi ,\
d\chi^\alpha \rightarrow g^\alpha ,\
d\psi^\alpha \rightarrow a^\alpha ,\
\nonumber
\end{eqnarray}
\begin{eqnarray}\label{b11}
d\pi^\alpha \rightarrow - p^\alpha ,\
dF \rightarrow \Xi ,\
d\varphi^\alpha \rightarrow f^\alpha\ .
\end{eqnarray}
We identify the grading of $d$ with the Grassmann grading $g$ of the
quantities
$\theta^\alpha$ ($g(d) = g(\theta^\alpha) = 1$), i.e., $g(dx^a) = g(x^a) +
1$.
The composition $\lambda \circ d$ of the maps $\lambda$ and $d$  renders
the even Hamiltonian (\ref{b8}) into the odd one (\ref{b5}):
$dH \stackrel{\lambda}{\rightarrow} \bar H $.

The inter-relation of the brackets (\ref{b7}), (\ref{b4}) and of the
corresponding to them Hamiltonians (\ref{b8}), (\ref{b5}) can be described
by the following scheme.  If we have a Hamilton system described in the
bracket $\{ A , B \}_{\epsilon }$ by means of the Hamiltonian
$\stackrel{\epsilon}{H}$
\begin{equation}\label{b12}
\stackrel{\cdot}{x^a} = \{x^a,
\stackrel{\epsilon}{H} \}_{\epsilon } =\ \stackrel{\epsilon}{\omega}^{ab}
 {{\partial {\stackrel{\epsilon}{H}}} \over \partial{x^b}} \ ,
\end{equation}
where $\epsilon$ $(\epsilon = 0,1)$ is the Grassmann parity of both the
bracket and the Hamiltonian, then the Hamilton equations for the phase
coordinates $x^a$ and the equations for their differentials $dx^a$,
obtaining by a differentiation of equations (\ref{b12}), can be reproduced
by the following bracket of the opposite Grassmann parity
\begin{eqnarray}\label{b13}
\{A,B\}_{\epsilon+1}
= A[ \stackrel{\leftarrow}{{\partial}\over
\partial{x^a} } \stackrel{\epsilon}{\omega}^{ab}
\stackrel{\rightarrow}{{\partial} \over \partial{(dx^b)}} \nn
+ (-1)^{g(a)+\epsilon}
\stackrel{\leftarrow}{{\partial} \over \partial{(dx^a)}}
\stackrel{\epsilon}{\omega}^{ab}
\stackrel{\rightarrow}{{\partial} \over \partial{x^b}} \nn
+ \stackrel{\leftarrow}{{\partial}\over \partial{(dx^a)} }
(d\stackrel{\epsilon}{\omega}^{ab})
\stackrel{\rightarrow}{{\partial} \over \partial{(dx^b)}}]B
\end{eqnarray}
with the help of the Hamiltonian $d{\stackrel{\epsilon}{H}}$
$(g(d{\stackrel{\epsilon}{H}}) = \epsilon + 1)$, that is,
\begin{eqnarray}
\stackrel{\cdot}{x^a} = \{x^a, \stackrel{\epsilon}{H} \}_{\epsilon } =\
 \{x^a, d{\stackrel{\epsilon}{H}} \}_{\epsilon + 1}\ ;
\nonumber
\end{eqnarray}
\begin{eqnarray}
\stackrel{\cdot}{dx^a} = d(\{x^a, \stackrel{\epsilon}{H} \}_{\epsilon }) =\
\{dx^a, d{\stackrel{\epsilon}{H}} \}_{\epsilon + 1}\ .
\nonumber
\end{eqnarray}
In connection with a similar scheme see also the paper \cite{n}.
Note also natural appearance of the odd bracket under exterior
differentiation of the equations of motion in \cite{vty}.

There is also an
interconnection between Lagrange's equations corresponding to
the Lagrangians of the different Grassmann parities $L$ (\ref{b9}) and
$\bar L$ (\ref{b2}), because the odd Lagrangian $\bar L$ (\ref{b2}) is
related by means of the redefinition $\lambda$ (\ref{b11}) with the
exterior differential $dL$ of the even Lagrangian (\ref{b9}).  Indeed,
Lagrange's equations for the Lagrangian $\stackrel{\epsilon}{L}(q^a,
\stackrel{\cdot}{q^a})$ with the Grassmann parity $\epsilon$ can be
written in the two equivalent forms
\begin{equation}\label{b14a}
{d \over dt} \left({\partial{\stackrel{\epsilon}{L}} \over
\partial{\stackrel{\cdot}{q^a}}}\right)
- {\partial{\stackrel{\epsilon}{L}} \over \partial{q^a}} = 0
\end{equation}
\begin{equation}\label{b14b}
{d \over dt} \left({\partial{(d{\stackrel{\epsilon}{L}})} \over
\partial{(\stackrel{\cdot}{dq^a}})}\right) -
{\partial{(d{\stackrel{\epsilon}{L}})} \over \partial{(dq^a)}} = 0\ ,
\end{equation}
while the equations obtained by taking the differential of (\ref{b14a})
have the form
\begin{equation}\label{b14c}
{d \over dt} \left({\partial{(d{\stackrel{\epsilon}{L}})} \over
\partial{\stackrel{\cdot}{q^a}}}\right) -
{\partial{(d{\stackrel{\epsilon}{L}})} \over \partial{q^a}} = 0\ .
\end{equation}
Equations (\ref{b14b}), (\ref{b14c}) can be considered as Lagrange's
equations for the system described in the configuration space
 $q^a$, $dq^a$ by the Lagrangian
$d{\stackrel{\epsilon}{L}}$ of the Grassmann parity $\epsilon + 1$.
If the Lagrangian $\stackrel{\epsilon}{L}$ has the
constraints $\varphi^i(q^a, p_a = {\partial{\stackrel{\epsilon}{L}}}
 / {\partial{\stackrel{\cdot}{q^a}}})$ satisfying the relations in the
bracket corresponding to $\stackrel{\epsilon}{L}$
\begin{eqnarray}
\{ \varphi^i, \varphi^k \}_{\epsilon} = f^{ik}\ ,
\nonumber
\end{eqnarray}
then the Lagrangian $d{\stackrel{\epsilon}{L}}$ will possess the
constraints $\varphi^i(q^a, p_a = {\partial(d{\stackrel{\epsilon}{L}})}
 /  {\partial({\stackrel{\cdot}{dq^a}})})$, coinciding with those
following from $\stackrel{\epsilon}{L}$, and $d\varphi^i$ obeying the
relations
\begin{eqnarray}
\{ \varphi^i, d\varphi^k \}_{\epsilon + 1} = f^{ik};\qquad
\{ \varphi^i, \varphi^k \}_{\epsilon + 1} = 0\ ,
\nonumber
\end{eqnarray}
which follow from the related with $d{\stackrel{\epsilon}{L}}$ bracket
expression (\ref{b13}) that in the case is without the last term in their
right-hand side (cf., e.g., equations (\ref{b10}) with (\ref{b3})).

Thus, it is shown that for the given formulation of the dynamics (either
in Hamilton's or in Lagrange's approach) with  the equations
of motion for the dynamical variables $x^a$ we can construct,
by using the exterior differential, such a formulation, having the
opposite Grassmann parity, that reproduces the former equations for $x^a$
and gives, besides, the equations for their differentials $dx^a$.

\section{QUANTUM ASPECT OF THE ODD BRACKET}

\subsection{Quantum representations of the odd Poisson bracket}

The procedure of the odd-bracket canonical quantization
given in \cite{vst,vs1} resides in splitting all the canonical
variables into two sets, in the division of all the functions
dependent on the canonical variables into classes, and in the
introduction of the quantum multiplication $*$, which is
either the common product or the bracket composition, in dependence on
what the classes the co-factors belong to. Under this, one of the classes
has to contain the normalized wave functions, and the result of
the multiplication $*$ for any quantity on the wave function $\Psi$ must
belong to the class containing $\Psi$. This procedure is the
generalization on the odd bracket case of the canonical quantization rules
for the usual Poisson bracket $\{\dots , \dots \}_{\rm Pois}$, which, for
example, in the coordinate representation for the canonical variables $q$
and $p$ are defined as
\begin{eqnarray}
q\ *\Psi(q) = q \Psi(q)\ ,
\nonumber
\end{eqnarray}
\begin{eqnarray}
p\ *\Psi(q) = i\hbar\{p, \Psi(q)\}_{\rm Pois} = - i\hbar
{\partial\Psi\over{\partial q}}\ ,
\nonumber
\end{eqnarray}
where $\Psi(q)$ is the normalized
wave function depending on the coordinate $q$.

In \cite{vst,vs1} two nonequivalent odd-bracket quantum
representations for the canonical variables were obtained by using two
different ways of the function division. But these ways do not exhaust all
the possibilities. In \cite{s2} a more general way of the
division is proposed, which contains as the limiting cases the ones given
in \cite{vst,vs1}.

Let us build quantum representations for an arbitrary graded bracket
under its canonical quantization. To this end, all canonical variables are
split into two equal in the number sets, so that none of them should
contain the pairs of canonical conjugates. Note that to make such
a splitting possible for the even bracket (1), the transition has to be
done from the real canonical self-conjugate odd variables to some pairs of
odd variables, which simultaneously are complex and canonical conjugate
to each other.  Composing from the integer degrees of the variables from
the one set (we call it the first set) the monomials of the odd $2s+1$
and even $2s$ uniformity degrees and multiplying them by the arbitrary
functions dependent on the variables from the other (second) set, we thus
divide all the functions of the canonical variables into the classes
designated as $\stackrel{\epsilon}{O}_s$ and $\stackrel{\epsilon}{E}_s$,
respectively.  For instance, in the general case the odd-bracket canonical
variables can be split, so that the first set would contain the even $y_i$
($i = 1,\dots,n\le N$) and odd $\eta ^{n+\alpha}$ ($\alpha = 1,\dots,N-n$)
variables, while the second set would involve the rest variables
\cite{s2}.  Then the classes of the functions obtained under this
splitting have the form
\begin{eqnarray}
\stackrel{1}{O}_s =
\left(y_i,\eta^{n+\alpha}\right)^{2s+1}f\left(\eta^i,y_{n+\alpha}\right)\ ;
\nonumber
\end{eqnarray}
\begin{eqnarray}
\stackrel{1}{E}_s =
\left(y_i,\eta^{n+\alpha}\right)^{2s}f\left(\eta^i,y_{n+\alpha}\right)\ ,
\nonumber
\end{eqnarray}
where the factors before the arbitrary function
$f\left(\eta^i,y_{n+\alpha}\right)$ denote the monomials having the
uniformity degrees indicated in the exponents. These classes satisfy the
corresponding bracket relations
\begin{eqnarray}
\{\stackrel{\epsilon}{O}_s,\stackrel{\epsilon}{O}_{s'}\}_{\epsilon } =
\stackrel{\epsilon}{O}_{s+s'}\ ;\qquad
\{\stackrel{\epsilon}{O}_s,\stackrel{\epsilon}{E}_{s'}\}_{\epsilon } =
\stackrel{\epsilon}{E}_{s+s'}\ ;
\nonumber
\end{eqnarray}
\begin{equation}\label{16}
\{\stackrel{\epsilon}{E}_s,\stackrel{\epsilon}{E}_{s'}\}_{\epsilon } =
\stackrel{\epsilon}{O}_{s+s'-1}\ ,
\end{equation}
and the relations of the ordinary Grassmann multiplication
\begin{eqnarray}
\stackrel{\epsilon}{O}_s\cdot\stackrel{\epsilon}{O}_{s'} =
\stackrel{\epsilon}{E}_{s+s'+1}\ ;\qquad
\stackrel{\epsilon}{O}_s\cdot\stackrel{\epsilon}{E}_{s'} =
\stackrel{\epsilon}{O}_{s+s'}\ ;
\nonumber
\end{eqnarray}
\begin{equation}\label{17}
\stackrel{\epsilon}{E}_s\cdot\stackrel{\epsilon}{E}_{s'} =
\stackrel{\epsilon}{E}_{s+s'}\ .
\end{equation}
It follows from (\ref{16}), (\ref{17}), that $\stackrel{\epsilon}{O} =
\{\stackrel{\epsilon}{O}_s\}$ and $\stackrel{\epsilon}{E} =
\{\stackrel{\epsilon}{E}_s\}$ form a superalgebra with respect to the
addition and the quantum multiplication $\qmul$ ($\epsilon = 0,1$) defined
for the corresponding bracket as
\begin{eqnarray}
\stackrel{\epsilon}{O}'\ \qmul\ \stackrel{\epsilon}{O}'' =
\{\stackrel{\epsilon}{O}',\stackrel{\epsilon}{O}''\}_{\epsilon }\in\
{\stackrel{\epsilon}{O}}\ ;
\nonumber
\end{eqnarray}
\begin{eqnarray}
\stackrel{\epsilon}{O}'\ \qmul\ \stackrel{\epsilon}{E}'' =
\{\stackrel{\epsilon}{O}',\stackrel{\epsilon}{E}''\}_{\epsilon }\in\
{\stackrel{\epsilon}{E}}\ ;
\nonumber
\end{eqnarray}
\begin{equation}\label{18}
\stackrel{\epsilon}{E}'\ \qmul\ \stackrel{\epsilon}{E}'' =
\stackrel{\epsilon}{E}'\cdot\stackrel{\epsilon}{E}''\in\
{\stackrel{\epsilon}{E}}\ ,
\end{equation}
where $\stackrel{\epsilon}{O}',\stackrel{\epsilon}{O}''\in
{\stackrel{\epsilon}{O}}$  and
$\stackrel{\epsilon}{E}', \stackrel{\epsilon}{E}''\in
{\stackrel{\epsilon}{E}}$. Note, that the classes
$\stackrel{\epsilon}{O}_0$ and $\stackrel{\epsilon}{E}_0$ form
the sub-superalgebra. In terms of the quantum grading $q_{\epsilon }(A)$
of any quantity $A$
\begin{eqnarray}
q_{\epsilon }(A) = \cases{g_{\epsilon
}(A),&for  $A\in{\stackrel{\epsilon}{O}}$;\cr g(A),&for
$A\in{\stackrel{\epsilon}{E}}$,\cr}
\nonumber
\end{eqnarray}
introduced for the appropriate
bracket, the grading and symmetry properties of the quantum multiplication
$\qmul$ , arising from the corresponding properties for the bracket
(\ref{3a}), (\ref{3b}) and Grassmann composition of any two quantities $A$
and $B$, are uniformly written as
\begin{equation}\label{19a}
q_{\epsilon }(A\ \qmul\ B) = q_{\epsilon }(A) +
q_{\epsilon }(B)\ ,
\end{equation}
\begin{equation}\label{19b}
\stackrel{\epsilon}{O}'\ \qmul\
\stackrel{\epsilon}{O}'' = -(-1) ^{q_\epsilon (\stackrel{\epsilon}{O}')
q_\epsilon (\stackrel{\epsilon}{O}'')}
\stackrel{\epsilon}{O}''\ \qmul\ \stackrel{\epsilon}{O}'\ ,
\end{equation}
\begin{equation}\label{19c}
\stackrel{\epsilon}{E}'\ \qmul\ \stackrel{\epsilon}{E}'' =
(-1) ^{q_\epsilon (\stackrel{\epsilon}{E}')
q_\epsilon (\stackrel{\epsilon}{E}'')}
\stackrel{\epsilon}{E}''\ \qmul\ \stackrel{\epsilon}{E}'\ .
\end{equation}

With the use of the quantum multiplication $\qmul\ $ and the quantum
grading $q_{\epsilon }$ , let us define for any two quantities $A, B$  the
quantum bracket ((anti)commutator) $[A, B \}_{\epsilon }$ (under its
action on the wave function $\Psi$ that is considered to belong to the
class $E$) in the form \cite{vst,vs1,s2}
\begin{eqnarray}\label{20}
[A, B \}_{\epsilon}\ \qmul\ \Psi  = A\ \qmul\ (B\ \qmul\ \Psi)\nn
-(-1)^{q_\epsilon (A)q_\epsilon (B)} B\ \qmul\ (A\ \qmul\ \Psi)\ .
\end{eqnarray}
If $A, B \in {\stackrel{\epsilon}{E}}$, then, due to (\ref{19c}), the
quantum bracket between them equals zero. In particular, the wave
functions are (anti)commutative. If $A$ or both of the quantities $A$ and
$B$ belong to the class ${\stackrel{\epsilon}{O}}$, then in the first
case, due to the Leibnitz rule (\ref{3d}), and in the second one, because
of the Jacobi identities (\ref{3c}), the relation follows from the
definitions (\ref{18}) and (\ref{20})
\begin{eqnarray}
[A , B \}_{\epsilon }\ \qmul\ \Psi  = \{A , B\}_{\epsilon }\ \qmul\ \Psi
= (A\ \qmul\ B)\ \qmul\ \Psi\ ,
\nonumber
\end{eqnarray}
that establishes the connection between the classical and quantum brackets
of the corresponding Grassmann parity. Note, that the
quantization procedure also admits the reduction to $O_{o}\cup E_{o}$.

The grading $q_\epsilon$ determines the symmetry properties of the quantum
bracket (\ref{20}). Under above-mentioned splitting of the odd-bracket
canonical variables into two sets, the grading $q_1$ equals unity for the
variables $y_i \in\ \stackrel{1}{O}$, $\eta^i \in\ \stackrel{1}{E}$ ($i =
1,\dots,n\le N$) and equal to zero for the rest canonical variables
$y_{n+\alpha} \in\ \stackrel{1}{E}$, $\eta^{n+\alpha} \in\
\stackrel{1}{O}$ ($\alpha = 1,\dots,N-n$). Therefore, in this case
the quantum odd bracket is represented with the anticommutators between
the quantities $y_i, \eta^i$ and with the commutators for the remaining
relations of the canonical variables.  If the roles of the first and the
second sets of the canonical variables change, then the quantum bracket
is represented with the anticommutators between $y_{n+\alpha},
\eta^{n+\alpha}$ and with the commutators in the other relations.
In \cite{vst,vs1} the odd-bracket quantum representations were
obtained for the cases $n = 0,N$, containing, respectively, only
commutators or anticommutators.

\subsection{Quantization of the systems with the odd Poisson bracket}

As the simplest example of using of the odd-bracket quantum
representations under the quantization of the
classical systems based on the odd bracket \cite{s2}, let us consider the
one-dimensional supersymmetric oscillator, whose phase superspace $x^ A$
contains a pair of even $q, p$ and a pair of odd $\eta ^1, \eta ^2$  real
canonical coordinates.  In terms of more suitable complex coordinates $z =
(p - iq)/\sqrt2$, $\eta = (\eta ^1 - i\eta ^2)/ \sqrt2$ and their complex
conjugates $\bar z, \bar \eta$, the even bracket is written as
\begin{eqnarray}\label{21}
\{ A, B\}_{0} = iA [\stackrel{\leftarrow}{\partial}_{\bar z}
\stackrel{\rightarrow}{\partial}_z -
\stackrel{\leftarrow}{\partial}_z
\stackrel{\rightarrow}{\partial}_{\bar z} -
(\stackrel{\leftarrow}{\partial}_{\bar \eta}
\stackrel{\rightarrow}{\partial}_\eta\nn
+\stackrel{\leftarrow}{\partial}_\eta
\stackrel{\rightarrow}{\partial}_{\bar \eta})] B
\end{eqnarray}
and the even Hamiltonian $H$, the supercharges $Q_1$, $Q_2$ and
the fermionic charge $F$ have the forms
\begin{eqnarray}
H = z\bar z + {\bar \eta} \eta\ ;\qquad Q_1 = \bar{z} \eta + z \bar\eta\ ;
\nonumber
\end{eqnarray}
\begin{equation}\label{22}
Q_ 2 =i(\bar{z} \eta - z \bar\eta)\ ;\qquad F = \eta \bar\eta\ .
\end{equation}
The odd Hamiltonian $\bar H$ and the appropriate odd bracket, which
reproduce the same Hamilton equations of motion, as those resulting from
(\ref{21}) with the even Hamiltonian $H$ (\ref{22}), i.e., which satisfy
the condition (\ref{9}), can be taken as $\bar H = Q_1$ and
\begin{eqnarray}\label{23}
\{A , B \}_{1} = iA (\stackrel{\leftarrow}{\partial}_{\bar z}
\stackrel{\rightarrow}{\partial}_{\eta} -
\stackrel{\leftarrow}{\partial}_{\eta}
\stackrel{\rightarrow}{\partial}_{\bar z} +
\stackrel{\leftarrow}{\partial}_{\bar \eta}
\stackrel{\rightarrow}{\partial}_z\nn
-\stackrel{\leftarrow}{\partial}_z
\stackrel{\rightarrow}{\partial}_{\bar \eta}) B\ .
\end{eqnarray}
The complex variables have the advantage over the real
ones, because with their use the splitting of the
canonical variables into two sets $\bar z , \bar \eta$ and $z , \eta $
satisfies simultaneously the requirements necessary for the quantization
both of the brackets (\ref{21}), (\ref{23}).  Besides, any of the vector
fields $\stackrel{\epsilon}{X}_{A_i} = - i \{ A_i ,\dots \}_{\epsilon }$
for the quantities $\{A_i\} = (H, Q_1, Q_2, F)$, describing the dynamics
and the symmetry of the system under consideration, is split into the sum
of two differential operators dependent on either $\bar z, \bar \eta$ or
$z , \eta$ . For instance, from (\ref{21})-(\ref{23}) we have
\begin{equation}\label{24}
\stackrel{0}{X}_H = \stackrel{1}{X}_{\bar H} = z\partial_z +
\eta\partial_\eta - \bar z\partial_{\bar z} - \bar \eta\partial_{\bar
\eta}\ .
\end{equation}
The diagonalization does not take place in terms of the variables
$x^A = (q, p ; \eta^1, \eta^2)$.

In accordance with the above-mentioned splitting of the complex variables,
we can perform one of the two possible divisions all of the functions into
the classes, which are common for both of the brackets (\ref{21}),
(\ref{23}), playing a crucial role under their canonical quantization and
leading to the same quantum dynamics for the system under consideration.
If $\bar z, \bar \eta$ are attributed to the first set, then the
corresponding function division is
\begin{eqnarray}
\stackrel{\epsilon}{O}_s =
(\bar z \bar\eta)^{2s+1} f (z,\eta)\ ;\qquad \stackrel{\epsilon}{E}_s =
(\bar z \bar\eta)^{2s} f (z,\eta)\ .
\nonumber
\end{eqnarray}

If we restrict ourselves to the classes $O_o$  and $E_o$, then
$\Psi \in\ E_o$ and depends only on $z, \eta $  and $A_i \in\ O_o$.
According to the definition (\ref{18}), the results of the quantum
multiplications $\qodd$ and $\qeven$ of $z , \eta \in\ E_o$ and $\bar z,
\bar\eta \in\ O_o$ on the wave function $\Psi$ are
\begin{eqnarray}
z \ \qodd\ \Psi = z\ \qeven\ \Psi = z \ \cdot\ \Psi\ ;
\nonumber
\end{eqnarray}
\begin{eqnarray}
\bar\eta\ \qodd\ \Psi =\bar z\ \qeven\ \Psi  = \partial_z\Psi\ ;
\nonumber
\end{eqnarray}
\begin{eqnarray}
\eta\ \qodd\ \Psi = \eta\ \qeven\ \Psi  = \eta\ \cdot\ \Psi\ ;
\nonumber
\end{eqnarray}
\begin{eqnarray}\label{25}
\bar z\ \qodd\ \Psi  = - \bar \eta\ \qeven\ \Psi  = \partial_{\eta}\Psi\ .
\end{eqnarray}
The positive definite scalar product of the
wave functions $\Psi_1(z,\eta)$ and $\Psi_2(z,\eta)$ can be determined in
the form \cite{ber2}
\begin{eqnarray}\label{26}
(\Psi_1, \Psi_2) = {1\over \pi} \int \exp
[-(\mid z\mid ^2 + \bar \theta \eta )]\ \Psi_1(z,\eta)\nn
\times[\Psi_2(z,\theta)]^+ d\bar\theta\,d\eta\,d(Re z)\,d(Im z)\ ,
\end{eqnarray}
where $\theta$ is the auxiliary complex Grassmann quantity
anticommuting with $\eta$, and the integration over the real and imaginary
components of $z$ is performed in the limits ($-\infty,\infty$). It is
easy to see that with respect to the scalar product (\ref{26}) the pairs
of the canonical variables, being Hermitian conjugated to each other under
the multiplication $\qodd$, are $z, \bar \eta$ and $\bar z, \eta$, but
under $\qeven$ are $z, \bar z$ and $\eta, -\bar \eta$.

In order to have the action of the Hamiltonian operator, obtained from
the system quantization, on the wave function, we need, as it is
well known, to replace the canonical variables in the classical
Hamiltonian by the respective operators or, which is the same, to
define their action with the help of the corresponding quantum
multiplication $\qmul$. In this connection, in view of (\ref{24}),
(\ref{25}), we see that the self-consistent quantum Hamilton operators in
the even and odd cases, being in agreement with the classical expressions
(\ref{22}) for the equivalent Hamiltonians $H$ and $\bar H$  and giving
the same result at the action on $\Psi(z, \eta) $ , will be respectively
\begin{equation}\label{27}
H\ \qeven\ \Psi = z\ \qeven\ (\bar z\ \qeven\ \Psi)\ -
\ \eta\ \qeven\ (\bar \eta\ \qeven\ \Psi)\ ;
\end{equation}
\begin{equation}\label{28}
\bar H\ \qodd\ \Psi  = z\ \qodd\ (\bar \eta\ \qodd\ \Psi)\ +
\ \eta\ \qodd\ (\bar z\ \qodd\ \Psi)\ .
\end{equation}
The Hamiltonians (\ref{27}), (\ref{28}) are Hermitian relative to the
scalar product (\ref{26}) and both, due to (\ref{25}), are reduced to the
Hamilton operator for the one-dimensional supersymmetric oscillator $H =
a^+a + b^+b$ expressed in terms of the creation and annihilation operators
for the bosons $a^{+} = z,\ a = \partial_z$ and fermions $b^{+} = \eta,\ b
= \partial _{\eta }$ respectively, in the Fock-Bargmann representation
(see, for example \cite{p}). The normalized with respect to (\ref{26})
eigenfunctions $\Psi_{k,n}(z,\eta)$ of the Hamiltonians (\ref{27}),
(\ref{28}), corresponding to energy eigenvalues $E_{k,n} = k + n$ ($k =
0,1; n = 0,1,\dots,\infty$) have the form
\begin{eqnarray}
\Psi_{k,n}(z,\eta) = {1\over\sqrt{n!}} \left(\eta\ \qmul\ \right)^{k}\
\left(z\ \qmul\ \right)^{n}1\ .
\nonumber
\end{eqnarray}
Note, that another equivalent representation of the quantum supersymmetric
oscillator can be obtained, if the canonical variables $z, \eta$ are
chosen as the first set.
Let us also note that the consideration described above can be extended to
the quantization of a set non-interacting supersymmetric oscillators by
supplement all the canonical variables $z, \bar z, \eta, \bar \eta$ with
the index $i$ $(i = 1,\dots,N)$ over which a summation have to be
performed in the bilinear combinations of the variables in all the
formulas of this subsection.

Thus, we have demonstrated that
the use of the quantum representations found for the odd bracket \cite{s2}
leads to the self-consistent quantization of the classical Hamilton
systems based on this bracket. We should apparently expect that these
representations are also applicable for the quantization of more
complicated classical systems with the odd bracket.

\subsection{Composite spinor structure of space--time}

The odd--bracket quantum representation with $n=N$ for the canonical
variables $y_{\alpha i}$, $\eta^{\beta k}$
\begin{eqnarray}
\{y_{\alpha i}, \eta^{\beta k}\}_1=\delta_\alpha^\beta\delta_i^k\ ,
\nonumber
\end{eqnarray}
\begin{equation}\label{c1}
\{y_{\alpha i}, y_{\beta k}\}_1=\{\eta^{\alpha i}, \eta^{\beta k}\}_1=0
\end{equation}
with the spinor indices $\alpha, \beta$ of the space--time symmetry and
with the indices $i, k$ of the internal symmetry can be applied for the
realization of the idea of a composite spinor structure of space--time
\cite{vst}.  Here we illustrate the realization of this idea on the
example of superspace of the orthosymplectic groups $OSp(N,2k)$ with
$k=1,2$, which are $SO(N)$-extended supergeneralizations of the de Sitter
groups $O(k,k+1)$, which admit real spinor representations. For the
quantum representation with $n=N$ the division of functions of the
canonical variables has the form
\begin{eqnarray}
\stackrel{1}{O}_s =
y^{2s+1} f (\eta)\ ;\qquad \stackrel{1}{E}_s =
y^{2s} f (\eta)\ .
\nonumber
\end{eqnarray}
The superspace coordinates $z^a=(x^\mu,\theta^{\alpha i})$ and the wave
function $\Psi(z)$ will be constructed from quantities belonging to
$\stackrel{1}{E}_0$, while the dynamical variables describing the symmetry
of the coordinates $z^a$ will be constructed from the quantities of the
class $\stackrel{1}{O}_0$.

Using the relations (\ref{c1}) and (\ref{3d}), we can convince ourselves
that the generators of supertranslations (here we follow the notations of
\cite{vst})
\begin{equation}\label{c2}
Q_{\alpha i}=y_{\alpha i}-i\kappa(\bar\eta_iy_k)\eta^k_\alpha
\end{equation}
together with the generators
\begin{equation}\label{c3}
L_{\alpha\beta}=-2iy_{i(\alpha}\eta_{\beta)}^i
\end{equation}
of the group $Sp(2k,R)$ which henceforth plays the role of the group of
the space--time symmetry, and the generators
\begin{equation}\label{c4}
I_{ik}=-3i(\bar y_{[i}\eta_{k]})
\end{equation}
of the internal--symmetry group $SO(N)$ form, with respect to the odd
bracket (\ref{c1}), the superalgebra of the group $OSp(N,2k)$. In the
definition (\ref{c2}) for the supertranslation generators we introduced a
dimensional parameter $\kappa$ with the meaning of the inverse radius of
curvature of the de Sitter space. Letting $\kappa$ tend to zero, we go
over to superalgebras of $N$-extended Poincar\'e supergroups: a
two--dimensional superalgebra when $k=1$, and a four--dimensional
superalgebra when $k=2$.

We now show how, with the aid of the representations
(\ref{c2})--(\ref{c4}) for the generators of the group $OSp(N,2k)$
$(k=1,2)$, it is possible to construct in terms of the quantities
$y_{\alpha i}$ and $\eta^{\beta k}$ the usual coordinates $x^\mu$ and
$\theta^{\alpha i}$ of de Sitter superspaces. It follows from the
relations (\ref{c1}), (\ref{3d}) and (\ref{c2}) that the spinors
$\eta^{\alpha i}$ and $y_{\alpha i}$ are transformed by the generators of
supertranslations as:
\begin{eqnarray}\label{c5}
\delta\eta^{\alpha i}=\{(\bar aQ), \eta^{\alpha i}\}_1=a^{\alpha i}
+i\kappa(\bar a^k\eta_i)\eta^\alpha_k,
\end{eqnarray}
\begin{eqnarray}\label{c6}
\delta y_{\alpha i}=\{(\bar aQ), y_{\alpha i}\}_1=
-i\kappa[a_\alpha^k(\bar\eta_ky_i)\nn
+(\bar a_i\eta_k)y_\alpha^k],
\end{eqnarray}
where $a^{\alpha i}$ are anticommuting parameters. The relation (\ref{c5})
is the transformation law under supertranslations for the coordinates
$\eta^{\alpha i}$ of the homogeneous space $OSp(N,2k)/SO(N)\otimes
Sp(2k,R)$, and these coordinates can be identified with the conventionally
used Grassmann variables $\theta^{\alpha i}$:
\begin{equation}\label{c7}
\theta^{\alpha i}=\eta^{\alpha i},
\end{equation}
whereas (\ref{c6}) can be represented in the form of a transformation of a
1--form field specified on this homogeneous space:
\begin{eqnarray}
y'_{\alpha i}={\partial\eta^{\beta k}\over\partial\eta'^{\alpha i}}
y_{\beta k}.
\nonumber
\end{eqnarray}
We shall transit from the variables $y_{\alpha i}$ to variables
$\tilde y_{\alpha i}(y,\eta)$ that are linear in $y_{\alpha i}$ and have
the following transformation law with respect to generators $Q_{\alpha i}$
\begin{equation}\label{c8}
\delta\tilde y_{\alpha i}=\{(\bar aQ), \tilde y_{\alpha i}\}_1
=i(\lambda_{\alpha\beta}\tilde y^\beta_i
+\lambda_{ik}\tilde y^k_\alpha),
\end{equation}
where $\lambda_{\alpha\beta}$ is a symmetric matrix and $\lambda_{ik}$ is
an antisymmetric one, both of which depend on the parameters
$a^{\alpha i}$ and the coordinates $\eta^{\alpha i}$. The form of the
functions $\tilde y_{\alpha i}(y,\eta)$ and the matrices
$\lambda_{\alpha\beta}(a,\eta)$ and $\lambda_{ik}(a,\eta)$ depends on the
values of $N$ and $k$. For the following analysis we shall need only the
fact of the existence of such variables $\tilde y_{\alpha i}$ and the
structure of their transformation law (\ref{c8}), which determines in
infinitesimal form that representation of the group $OSp(N,2k)$ which is
induced by the representation of its even subgroup $SO(N)\otimes Sp(2k,R)$
that is the vector representation with respect to $SO(N)$ and the defining
representation with respect to the groups $Sp(2k,R)$.

For the construction of the space--time coordinates $x^\mu$ from the
quantities $\tilde y_{\alpha i}$ we define supertransformations of $x^\mu$
using that representation of the group $OSp(N,2k)$ which is induced by the
vector representation of the de Sitter group $SO(k,k+1)$. With respect to
the internal--symmetry group $SO(N)$ the inducing representation is taken
to be a scalar. By taking into account the transformation law (\ref{c8})
for the quantities $\tilde y_{\alpha i}$, we shall construct from them a
real vector $V^A$ with respect to the group $SO(k,k+1)$. The simplest
expression for $V^A$, containing the lowest powers of
$\tilde y_{\alpha i}$, has the form
\begin{equation}\label{c9}
V^A=(\bar{\tilde y}_i\Gamma^A\tilde y_k)g^{ik}
\end{equation}
for the case $k=1$, and
\begin{equation}\label{c10}
V^A=(\bar{\tilde y}_i\Gamma^A\tilde y_k)(\bar(\tilde y)^i\tilde y^k)
\end{equation}
for the case $k=2$.

One should note the remarkable fact that in the case $k=2$, because of the
antisymmetry of the matrices $C^{-1}\Gamma$, the vector $V^A$ can be
constructed only if we introduce an internal--symmetry group $SO(N)$ with
$N\geq2$, whereas for $k=1$ these matrices are symmetric and $V^A$ can be
constructed in the case $N=1$. However, in this case $V^AV_A=0$, i.e.,
such a vector is isotropic.

We shall define the coordinates $x^\mu$ for both values of $k$ in terms of
the corresponding expressions for $V^A$ in the form
\begin{equation}\label{c11}
x^\mu={V^\mu\over{\kappa V^5}}
\end{equation}
with the condition that the quantity $V^2=V^AV_A$, which is invariant of
$OSp(N,2k)$, has the sign which, for each value of $k$, leads to the
correct signature of the space--time metric, i.e., $V^2$ should be
positive when $k=1$ and negative when $k=2$. In the cases that we have
considered we did not succeed in satisfying these conditions for compact
internal--symmetry groups $SO(N)$. By going over to the non-compact groups
$SO(n,m)$ with $n+m=N$ it is possible, albeit not for all values of $N$,
to achieve the correct signature of the space--time metric. It is possible
that this circumstance is not accidental and is connected with the
well-known fact that in $N\geq4$ extended supergravities the scalar fields
form homogeneous non-compact spaces \cite{cj}.

We note one feature that appears to us to be curious. If besides the
coordinates $x^\mu$ and $\theta^{\alpha i}$ we construct composite
coordinates $u^{ik}$, corresponding to the internal--symmetry generators
$I_{ik}$, then the quadratic equation
\begin{eqnarray}
2k+{{N(N-1)}\over 2}=2^kN,
\nonumber
\end{eqnarray}
reflecting the equality of the numbers of degrees of freedom of the
coordinates $x^\mu$ and $u^{ik}$ and of their constituent spinor
$y_{\alpha i}$, has the solutions $N=1,4$ for $k=1$ and $N=1,8$ for $k=2$.
We must discard the solutions with $N=1$, since for these it is impossible
to construct $x^\mu$ with the appropriate signature and non-singular
character. The remaining solutions imply that for the composite spinor
structure of the superspace the value of $N$ is subject to a restriction
that coincides with the restriction that follows from consideration of the
maximal multiplet of extended supergravity.

Using the transformation law (\ref{c8}) for $\tilde y_{\alpha i}$ and the
expressions (\ref{c9})--(\ref{c11}) for the coordinates $x^\mu$, we can
obtain the transformation of $x^\mu$ under supertranslations; the form of
this transformation is determined by the structure of the matrices
$\lambda_{\alpha\beta}$, i.e., depends on the values of $N$ and $k$. As an
example we shall give the transformation of $x^\mu$ for the case of the
group $OSp(1,1;2)$:
\begin{eqnarray}\label{c12}
\delta x^\mu=\{(\bar aQ), x^\mu\}_1=-i(\bar a^k\gamma^\mu\eta_k)\nn
+2i\kappa(\bar a^k\sigma^{\mu\nu}\eta_k)x_\nu
+i\kappa^2(\bar a^k\gamma^\nu\eta_k)x_\nu x^\mu,
\end{eqnarray}
where the term independent of $\kappa$ for the other values of $N$ and $k$
will be the same as in (\ref{c12}), whereas the terms containing $\kappa$
can be different.

From the relations (\ref{c9})--(\ref{c11}) we obtain the following bracket
relations for angular momentum
\begin{eqnarray}
M_{\mu\nu}=-i(\bar y^k\sigma_{\mu\nu}\eta_k),
\nonumber
\end{eqnarray}
the momentum
\begin{eqnarray}
P_\mu={i\over 2}\kappa(\bar y^k\gamma_\mu\eta_k)
\nonumber
\end{eqnarray}
and the generators $I_{ik}$ with coordinates $x^\mu$ and $\theta^{\alpha
i}$:
\begin{eqnarray}
\{P_\mu, x^\nu\}_1=-i(\delta^\nu_\mu-\kappa^2x_\mu x^\nu),
\nonumber
\end{eqnarray}
\begin{eqnarray}
\{M_{\mu\nu}, x^\rho\}_1=-i(x_\mu\delta^\rho_\nu-x_\nu\delta^\rho_\mu),
\nonumber
\end{eqnarray}
\begin{eqnarray}
\{I_{ik}, x^\mu\}_1=0,
\nonumber
\end{eqnarray}
\begin{eqnarray}
\{P_\mu, \theta^{\alpha i}\}_1=
{i\over2}\kappa(\bar\theta^i\gamma_\mu)^\alpha,
\nonumber
\end{eqnarray}
\begin{eqnarray}
\{M_{\mu\nu}, \theta^{\alpha i}\}_1=
-i(\bar\theta^i\sigma_{\mu\nu})^\alpha,
\nonumber
\end{eqnarray}
\begin{eqnarray}\label{c13}
\{I_{ik}, \theta^{\alpha l}\}_1=
i(\delta^l_i\theta^\alpha_k-\delta^l_k\theta^\alpha_i),
\end{eqnarray}
where for each value of $k$ it is necessary to take the matrices
$\gamma_\mu$ and $\sigma_{\mu\nu}$ corresponding to it.

As $\kappa$ tends to zero the supergroups $OSp(n,m;2k)$ go over into
extended Poincar\'e supergroups, and at the same time the relations
(\ref{c5}), (\ref{c12}) and (\ref{c13}), with allowance for the
relationship between the classical and quantum odd brackets, reproduce, on
the composite coordinates $z^a=(x^\mu, \theta^{\alpha i})$ and dynamical
quantities $Q_{\alpha i}$, $P_\mu$, $M_{\mu\nu}$ and $I_{ik}$, describing
their symmetry, the algebra of the conventional canonical quantization of
the coordinates and dynamical quantities.

\section{ALGEBRAIC ASPECT OF THE ODD BRACKET}

\subsection{Linear odd bracket on Grassmann algebra}

In this section we show that the linear odd bracket can be realized solely 
on the Grassmann variables \cite{s3}. It will be also shown that with such 
a bracket, that corresponds to a semi-simple Lie group, both a 
Grassmann--odd Casimir function and invariant (with respect to this group)
nilpotent differential operators of the first, second and third
orders are naturally related and enter into a finite-dimensional Lie
superalgebra. It will be also pointed out that some of the quantities,
forming this Lie superalgebra, can be treated as the BRST charge,
$\Delta$--operator and operator for the ghost number .

There is a well-known linear even Poisson bracket given
in terms of the commuting variables $X_\alpha$
\begin{equation}\label{d1}
\{X_\alpha, X_\beta \}_0 = {c_{\alpha\beta}}^\gamma X_\gamma,\
(\alpha,\beta,\gamma = 1,...,N),
\end{equation}
where constants  ${c_{\alpha\beta}}^\gamma$, because of the main
properties of the even Poisson bracket (\ref{3a})--(\ref{3d}),
are antisymmetric in the two lower indices
\begin{equation}\label{d2}
{c_{\alpha\beta}}^\gamma = - {c_{\beta\alpha}}^\gamma
\end{equation}
and obey the conditions
\begin{equation}\label{d3}
{c_{\alpha\lambda}}^\delta {c_{\beta\gamma}}^\lambda +
{c_{\beta\lambda}}^\delta {c_{\gamma\alpha}}^\lambda +
{c_{\gamma\lambda}}^\delta {c_{\alpha\beta}}^\lambda = 0\ .
\end{equation}
The linear even bracket (\ref{d1}) plays a very important
role in the theory of Lie groups, Lie algebras, their representations and
applications (see, for example, \cite{ber1,km}). The bracket (\ref{d1})
can be realized in a canonical even Poisson bracket
\begin{eqnarray}
\{ A , B \}_0 = A \sum_{\alpha = 1}^N
\left(\lpar_{q^\alpha}
\rpar_{p_\alpha} -
\lpar_{p_\alpha}
\rpar_{q^\alpha}\right) B
\nonumber
\end{eqnarray}
on the following bilinear functions of coordinates $q^\alpha$ and
momenta $p_\alpha$
\begin{eqnarray}
X_\alpha = {c_{\alpha\beta}}^\gamma q^\beta p_\gamma
\nonumber
\end{eqnarray}
if ${c_{\alpha\beta}}^\gamma$ satisfy the conditions (\ref{d2}), (\ref{d3})
for the structure constants of a Lie group.

As in the Lie algebra case, we can define a symmetric Cartan-Killing
tensor
\begin{equation}\label{d4}
g_{\alpha\beta} = g_{\beta\alpha} =
{c_{\alpha\gamma}}^\lambda {c_{\beta\lambda}}^\gamma
\end{equation}
and verify with the use of relations (\ref{d3}) an anti-symmetry property
of a tensor
\begin{equation}\label{d5}
c_{\alpha\beta\gamma} = {c_{\alpha\beta}}^\delta g_{\delta\gamma} =
- c_{\alpha\gamma\beta}\ .
\end{equation}
By assuming that the Cartan-Killing metric tensor is non-degenerate
$det(g_{\alpha\beta}) \neq 0$ (this case corresponds to the semi-simple
Lie group), we can define an inverse tensor $g^{\alpha\beta}$
\begin{equation}\label{d6}
g^{\alpha\beta}g_{\beta\gamma} = \delta^\alpha_\gamma\ ,
\end{equation}
with the help of which we are able to build a quantity
\begin{eqnarray}
C = X_\alpha X_\beta g^{\alpha\beta}\ ,
\nonumber
\end{eqnarray}
that, in consequence of relation (\ref{d5}), is for the bracket (\ref{d1})
a Casimir function which annihilates the bracket (\ref{d1}) and is an
invariant of the Lie group with the structure constants
${c_{\alpha\beta}}^\gamma$ and the generators $T_\alpha$
\begin{eqnarray}
\{ X_\alpha, C \}_0 =
{c_{\alpha\beta}}^\gamma X_\gamma \partial_{X_\beta} C =
T_\alpha C = 0\ .
\nonumber
\end{eqnarray}

Now let us replace in expression (\ref{d1}) the commuting variables
$X_\alpha$ by Grassmann variables $\Theta_\alpha$ $(g(\Theta_\alpha) = 1)$.
Then we obtain a binary composition
\begin{equation}\label{d7}
\{\Theta_\alpha, \Theta_\beta \}_1 =
{c_{\alpha\beta}}^\gamma \Theta_\gamma\ ,
\end{equation}
which, due to relations (\ref{d2}) and (\ref{d3}), meets all the
properties (\ref{3a})--(\ref{3d}) of the odd Poisson brackets.
It is surprising enough that the odd bracket can be defined solely
in terms of the Grassmann variables as well as an even Martin bracket
\cite{mar}. On the following bilinear functions of canonical variables
commuting $q^\alpha$ and Grassmann $\theta_\alpha$
\begin{eqnarray}
\Theta_\alpha = {c_{\alpha\beta}}^\gamma q^\beta \theta_\gamma
\nonumber
\end{eqnarray}
a canonical odd Poisson bracket
\begin{eqnarray}
\{ A , B \}_1 = A \sum_{\alpha = 1}^N
\left(\lpar_{q^\alpha}
\rpar_{\theta_\alpha} -
\lpar_{\theta_\alpha}
\rpar_{q^\alpha}\right) B
\nonumber
\end{eqnarray}
is reduced to the bracket (\ref{d7}) providing that
${c_{\alpha\beta}}^\gamma$ obey the conditions (\ref{d2}), (\ref{d3}).

By the way, let us note that we can also construct only on the
Grassmann variables a non-linear odd Poisson bracket of the form
\begin{eqnarray}
\{\Theta_\alpha, \Theta_\beta \}_1 =
{c_{\alpha\beta}}^{\gamma\delta\lambda}
\Theta_\gamma \Theta_\delta \Theta_\lambda\ .
\nonumber
\end{eqnarray}
In order to satisfy the property (\ref{3b}), the constants
${c_{\alpha\beta}}^{\gamma\delta\lambda}$ have to be antisymmetric in the
below indices
\begin{eqnarray}
{c_{\alpha\beta}}^{\gamma\delta\lambda} =
- {c_{\beta\alpha}}^{\gamma\delta\lambda}
\nonumber
\end{eqnarray}
and, for the validity of the Jacobi identities (\ref{3c}), they must obey
the following conditions
\begin{eqnarray}
\sum_{(\alpha\beta\gamma)} {c_{\alpha\beta}}^{\lambda[\delta_1\delta_2}
{c_{\lambda\gamma}}^{\delta_3\delta_4\delta_5]} = 0\ ,
\nonumber
\end{eqnarray}
where square brackets $[...]$ mean an antisymmetrization of the indices in
them.

Returning to the linear odd bracket (\ref{d7}) notice, that
on functions $A, B$ of Grassmann variables $\Theta_\alpha$ this bracket
has the form
\begin{eqnarray}
\{ A , B \}_1 = A \lpar_{\Theta_\alpha}
{c_{\alpha\beta}}^\gamma \Theta_\gamma \rpar_{\Theta_\beta} B\ .
\nonumber
\end{eqnarray}
The bracket (\ref{d7}) can be either degenerate or non-degenerate in the
dependence on whether the matrix ${c_{\alpha\beta}}^\gamma \Theta_\gamma$
in the indices $\alpha, \beta$ is degenerate or not. Raising and lowering
of the indices $\alpha, \beta$, the non-degenerate
metric tensors (\ref{d4}), (\ref{d6}) relate with each other the adjoint
and co-adjoint representations which are equivalent for a semi-simple Lie
group
\begin{eqnarray}
\Theta^\alpha = g^{\alpha\beta} \Theta_\beta\ ,\qquad
\partial_{\Theta^\alpha} =
g_{\alpha\beta} \partial_{\Theta_\beta}\ .
\nonumber
\end{eqnarray}
Hereafter only the non-degenerate metric tensors (\ref{d6}) will be
considered.

By contracting the indices in a product of the Grassmann
variables with the upper indices
and of the successive Grassmann derivatives,
respectively, with the lower indices
in (\ref{d3}), we obtain the relations
\begin{equation}\label{d8}
\Theta^\alpha \Theta^\beta
( {c_{\alpha\beta}}^\lambda {c_{\lambda\gamma}}^\delta +
2 {c_{\gamma\alpha}}^\lambda {c_{\lambda\beta}}^\delta ) = 0\ ,
\end{equation}
\begin{equation}\label{d9}
\Theta^\alpha \Theta^\beta \Theta^\gamma
{c_{\alpha\beta}}^\lambda {c_{\lambda\gamma}}^\delta = 0\ ,
\end{equation}
\begin{equation}\label{d10}
( {c_{\alpha\beta}}^\lambda {c_{\lambda\gamma}}^\delta +
2 {c_{\gamma\alpha}}^\lambda {c_{\lambda\beta}}^\delta )
\partial_{\Theta_\alpha} \partial_{\Theta_\beta} = 0\ ,
\end{equation}
\begin{equation}\label{d11}
{c_{\alpha\beta}}^\lambda {c_{\lambda\gamma}}^\delta
\partial_{\Theta_\alpha} \partial_{\Theta_\beta}
\partial_{\Theta_\gamma} = 0\ ,
\end{equation}
which will be used later on many times. In particular, taking into account
relation (\ref{d9}), we can verify that the linear odd bracket (\ref{d7})
has the following Grassmann-odd nilpotent Casimir function
\begin{equation}\label{d12}
\Delta_{+3} = {1\over\sqrt{3!}} \Theta^\alpha \Theta^\beta \Theta^\gamma
c_{\alpha\beta\gamma},\qquad(\Delta_{+3})^2 = 0,
\end{equation}
which is an invariant of the Lie group
\begin{eqnarray}\label{d13}
\{ \Theta_\alpha, \Delta_{+3} \}_1 =
\Theta_\gamma {c_{\alpha\beta}}^\gamma
\partial_{\Theta_\beta} \Delta_{+3} =
S_\alpha \Delta_{+3}\nn = 0
\end{eqnarray}
with the generators $S_\alpha$ obeying
the Lie algebra permutation relations (below $[A, B] =
AB -BA$ and $\{ A, B \} = AB + BA$)
\begin{equation}\label{d14}
[ S_\alpha, S_\beta ] = {c_{\alpha\beta}}^\gamma S_\gamma\ .
\end{equation}

It is a well-known fact that, in contrast with the even Poisson bracket,
the non-degenerate  odd Poisson bracket has one Grassmann-odd nilpotent
differential $\Delta$-operator of the second order, in terms of which the
main equation has been formulated in the Batalin-Vilkovisky scheme
\cite{bv} for the quantization of gauge theories in
the Lagrangian approach. In a formulation of Hamiltonian dynamics
by means of the odd Poisson bracket with the help of a Grassmann-odd
Hamiltonian $\bar H$ $(g(\bar H) = 1)$
\cite{l,vpst,s} this
$\Delta$-operator plays also a very important role being used to
distinguish the Hamiltonian dynamical systems, for which the Liouville
theorem is valid $\Delta \bar H = 0$, from those ones,
for which this theorem takes no place $\Delta \bar H \neq 0$.

Now let us try to build the $\Delta$-operator for the linear
odd bracket (\ref{d7}). It is remarkable that, in contrast with the
canonical odd Poisson bracket having the only $\Delta$-operator of the
second order, we are able to construct at once three $\Delta$-like
Grassmann-odd nilpotent operators which are differential operators of
the first, second and third orders respectively
\begin{equation}\label{d15}
\Delta_{+1} = {1\over\sqrt{2}} \Theta^\alpha \Theta^\beta
{c_{\alpha\beta}}^\gamma \partial_{\Theta^\gamma},
\qquad (\Delta_{+1})^2 = 0;
\end{equation}
\begin{equation}\label{d16}
\Delta_{-1} = {1\over\sqrt{2}} \Theta_\gamma
{c_{\alpha\beta}}^\gamma
\partial_{\Theta_\alpha} \partial_{\Theta_\beta},
\qquad (\Delta_{-1})^2 = 0;
\end{equation}
\begin{equation}\label{d17}
\Delta_{-3} = {1\over\sqrt{3!}} c_{\alpha\beta\gamma}
\partial_{\Theta_\alpha} \partial_{\Theta_\beta} \partial_{\Theta_\gamma}
,\ \ \ \ (\Delta_{-3})^2 = 0.
\end{equation}
The nilpotency of the operators $\Delta_{+1}$ and $\Delta_{-1}$ is a
consequence of relations (\ref{d9}) and (\ref{d11}). The operator
$\Delta_{+1}$ is proportional to the second term in a BRST charge
\begin{eqnarray}
Q = \Theta^\alpha G_\alpha - {1\over 2} \Theta^\alpha \Theta^\beta
{c_{\alpha\beta}}^\gamma \partial_{\Theta^\gamma}\ ,
\nonumber
\end{eqnarray}
where $\Theta^\alpha$ and $\partial_{\Theta^\alpha}$ represent the
operators for ghosts and antighosts respectively. $Q$ itself will be
proportional to the operator $\Delta_{+1}$ if we take the
representation $S_\alpha$ (\ref{d13}) for group generators $G_\alpha$. The
operator $\Delta_{-1}$, related with the divergence of a vector field
$\{ \Theta_\alpha, A \}_1$
\begin{eqnarray}
\partial_{\Theta_\alpha} \{ \Theta_\alpha, A \}_1 =
\partial_{\Theta_\alpha} S_\alpha A =
- \sqrt{2} \Delta_{-1} A\ ,
\nonumber
\end{eqnarray}
is proportional to the true $\Delta$-operator for the bracket (\ref{d7}).

It is also interesting to reveal that these $\Delta$-like operators
together with the Casimir function $\Delta_{+3}$ (\ref{d12}) are closed
into the finite-dimensional Lie superalgebra, in which the anticommuting
relations between the quantities $\Delta_\lambda$ $(\lambda = -3, -1, +1,
+3 )$ (\ref{d12}), (\ref{d15})--(\ref{d17}) with the nonzero right-hand
side are
\begin{equation}\label{d18}
\{ \Delta_{-1}, \Delta_{+1} \} = Z\ ,
\end{equation}
\begin{equation}\label{d19}
\{ \Delta_{-3}, \Delta_{+3} \} = N - 3Z\ ,
\end{equation}
where
\begin{eqnarray}
N = - c^{\alpha\beta\gamma} c_{\alpha\beta\gamma}
\nonumber
\end{eqnarray}
is a number of values for the indices $\alpha, \beta, \gamma$
$(\alpha, \beta, \gamma = 1,...,N)$ and
\begin{equation}\label{d20}
Z = D - K
\end{equation}
is a central element of this superalgebra
\begin{equation}\label{d21}
[ Z, \Delta_\lambda ] = 0\ ,\qquad (\lambda = -3, -1, +1, +3)\ .
\end{equation}
In (\ref{d20})
\begin{equation}\label{d22}
D = \Theta^\alpha \partial_{\Theta^\alpha}
\end{equation}
is a "dilatation" operator for the Grassmann variables
$\Theta_\alpha$, which distinguishes the $\Delta_\lambda$-operators
with respect to their uniformity degrees in $\Theta$
\begin{equation}\label{d23}
[ D, \Delta_\lambda ] = \lambda
\Delta_\lambda\ ,\qquad (\lambda = -3, -1, +1, +3)
\end{equation}
and is in fact a representation for a ghost number operator,
and the quantity $K$ has the form
\begin{equation}\label{d24}
K = {1\over 2} \Theta^\alpha \Theta^\beta
{c_{\alpha\beta}}^\lambda c_{\lambda\gamma\delta}
\partial_{\Theta_\gamma} \partial_{\Theta_\delta}\ .
\end{equation}
The operator $Z$ is also a central element of the Lie superalgebra which
contains both the operators $\Delta_\lambda$ (\ref{d12}),
(\ref{d15})--(\ref{d17}), $Z$ (\ref{d20}) and the operator $D$ (\ref{d22})
\begin{equation}\label{d25}
[Z, D] = 0\ .
\end{equation}

We can add to this superalgebra the generators $S_\alpha$ (\ref{d13})
with the following commutation relations:
\begin{equation}\label{d26}
[ S_\alpha, \Delta_\lambda ] = 0,\
[ S_\alpha, Z ] = 0,\ [ S_\alpha, D ] = 0,
\end{equation}
which indicate that both the Casimir function $\Delta_{+3}$ and the
operators $\Delta_\lambda$ $(\lambda = -3, -1, +1 )$, $Z$ and $D$ are
invariants of the Lie group with the generators $S_\alpha$.
In order to prove the permutation relations for the Lie superalgebra
(\ref{d12})--(\ref{d26}), we have to use relations (\ref{d8})--(\ref{d11}).
Note that the central element $Z$ (\ref{d20}) coincides with the
expression for a quadratic Casimir operator of the Lie algebra (\ref{d14})
for the generators $S_\alpha$ given in the representation (\ref{d13})
\begin{equation}\label{d27}
S_\alpha S_\beta g^{\alpha\beta} = Z\ .
\end{equation}

Thus, we see that both the even and odd linear Poisson brackets
are internally inherent in the Lie group with the structure constants
subjected to conditions (\ref{d2}) and (\ref{d3}).  However, only for the
linear odd Poisson bracket realized in terms of the Grassmann variables
and only in the case when this bracket corresponds to the semi-simple Lie
group, there exists the Lie superalgebra (\ref{d12})--(\ref{d26}) for the
$\Delta$-like operators of this bracket.

Note that in the case of the degenerate Cartan-Killing
metric tensor (\ref{d4}), relation (\ref{d5}) remains valid and we can
construct only two $\Delta$-like Grassmann-odd nilpotent operators:
$\Delta_{-1}$ (\ref{d16}) and $\Delta_{-3}$ (\ref{d17}), which satisfy the
trivial anticommuting relation
\begin{eqnarray}
\{ \Delta_{-1}, \Delta_{-3} \} = 0\ .
\nonumber
\end{eqnarray}
Note also that anticommuting relations
\begin{eqnarray}
\{ \stackrel{i}\Delta_{-1}, \stackrel{k}\Delta_{-1} \}=
-2 \stackrel{\{i}{{c_{\alpha\beta}}^\lambda}
\ \stackrel{k \}}{{c_{\lambda\gamma}}^\delta} \Theta_\delta
\partial_{\Theta_\alpha}\partial_{\Theta_\beta}\partial_{\Theta_\gamma}
\nonumber
\end{eqnarray}
for the operators
\begin{eqnarray}
\stackrel{i}\Delta_{-1} = {1\over\sqrt{2}} \Theta_\gamma
\stackrel{i}{{c_{\alpha\beta}}^\gamma}
\partial_{\Theta_\alpha} \partial_{\Theta_\beta}
\ ,\qquad (\stackrel{i}\Delta_{-1})^2 = 0\ ,
\nonumber
\end{eqnarray}
corresponding to the Lie algebras with structure constants
$\stackrel{i}{{c_{\alpha\beta}}^\gamma}$ $(i = 1,2)$,
vanish provided that $\stackrel{i}{{c_{\alpha\beta}}^\gamma}$ satisfy
compatibility conditions \cite{blt}
\begin{eqnarray}
\sum_{(\alpha\beta\gamma)} \stackrel{\{i}{{c_{\alpha\beta}}^\lambda}
\ \stackrel{k \}}{{c_{\lambda\gamma}}^\delta} = 0\ ,
\nonumber
\end{eqnarray}
where $\{ ik \}$ denotes a symmetrization of the indices $i$ and $k$.

The Lie superalgebra (\ref{d12})--(\ref{d26}), naturally connected with
the linear odd Poisson bracket (\ref{d7}), may be useful for the
subsequent development of the Batalin-Vilkovisky formalism for the
quantization of gauge theories. Let us note that this superalgebra can also
be used in the theory of representations of the semi-simple Lie groups.

\section*{Acknowledgments}

I am sincerely grateful to the Organizing Committee of this Symposium and,
especially, to Keith Olive and Mikhail Shifman for the financial support
necessary for my participation in this meeting and for the opportunity to
deliver this report.

\end{document}